\documentclass[12pt,preprint]{aastex}
 
\begin{document}
 
\title{The Transition between Nonorthogonal Polarization Modes in PSR B2016+28 
       at 1404 MHz}
\author{Mark M. McKinnon}
\affil{National Radio Astronomy Observatory\altaffilmark{1}\altaffiltext{1}
{The National Radio Astronomy Observatory is a facility of the National Science
Foundation operated under cooperative agreement by Associated Universities, Inc.},
Socorro, NM\ \ 87801\ \ USA}
\email{mmckinno@nrao.edu} 
\begin{abstract}

  Polarization observations of the radio emission from PSR B2016+28 at 1404 
MHz reveal properties that are consistent with two, very different, 
interpretations of the pulsar's viewing geometry. The pulsar's average 
polarization properties show a rapid change in position angle (PA) near the 
pulse center, suggesting that the observer's sightline nearly intersects the 
star's magnetic pole. But single pulse, polarization observations of the pulsar 
show nearly orthogonal modes of polarization following relatively flat and 
parallel PA trajectories across the pulse, suggesting that the sightline is 
far from the pole. Additionally, PA histograms reveal a \lq\lq modal 
connecting bridge", of unknown origin, joining the modal PA trajectories over 
much of the pulse and following the rapid PA change shown in the average 
data. The nonorthogonality of polarization modes is incorporated in a 
statistical model of radio polarization to account for the deviations from 
mode orthogonality that are observed in the pulsar. The model is used to 
interpret the rapid PA change and modal connecting bridge as a 
longitudinally-resolved transition between modes of nonorthogonal polarization. 
Thus, the modal PA trajectories are argued to reflect the pulsar's true viewing 
geometry. This interpretation is consistent with the pulsar's morphological 
classification, preserves the Radhakrishnan \& Cooke model of pulsar radio 
emission, and avoids the complication that the modal connecting bridge might 
be produced by some other emission mechanism. The statistical model's ability 
to simulate the rich variety of polarization properties observed in the 
emission lends additional support to the model's applicability and its 
underlying assumption that the polarization modes occur simultaneously.

\end{abstract}


\section{INTRODUCTION}
\label{sec:intro}

  All modern polar cap theories of pulsar radio emission are based upon
the hypothesis of Radhakrishnan \& Cooke (1969, hereafter RC) that the 
emission arises from charged particles streaming along open magnetic field 
lines above the pulsar's polar cap. The observed polarization due to this 
particle motion is the instantaneous projection of the magnetic field lines 
in the emission region on the plane of the sky. If the magnetic field 
structure is dipolar, the position angle (PA) of the resulting linear 
polarization vector varies in an S-shaped pattern with pulse longitude as 
the pulsar's rotation alters the field-line projection with respect to an 
observer's sightline. Since the polarization vector's orientation is 
determined solely by the dipolar field structure and the viewing geometry, 
the observed PAs are essentially frequency-independent. The astonishing ability 
of the RC model to explain both the PA patterns and the large fractional 
linear polarization that are actually observed is a testament to the model's 
success and popularity for over three decades.

  The 1404-MHz polarization observations of individual pulses from PSR 
B2016+28 by Stinebring et al. (1984, hereafter SCRWB) offer two very 
different interpretations of the pulsars' viewing geometry, thereby 
challenging the validity of the RC model (SCRWB). The longitude-dependent PA
histograms presented by SCRWB reveal two, nearly flat, PA traces separated by 
about $90^\circ$ across most of the pulse with a connecting bridge between the 
traces near the pulse center. The two PA traces most certainly arise from the 
well-known orthogonal modes of polarization (e.g. Manchester, Taylor, \& 
Huguenin 1975, hereafter MTH; Backer, Rankin, \& Campbell 1976, hereafter BRC; 
Cordes, Rankin, \& Backer 1978; Backer \& Rankin 1980, hereafter BR; SCRWB), 
and their flat trajectories suggest a viewing geometry where the observer's 
sightline makes a large angle with respect to the star's magnetic pole. But 
when the SCRWB data are averaged over many rotations of the star, the resulting 
PAs vary in the classic S-shaped pattern with a steep slope near the pulse 
center (Fig.~\ref{fig:profile}), suggesting that the sightline nearly intersects 
the magnetic pole. The steep PA sweep observed in the averaged data is embedded 
in the modal connecting bridge shown in the single pulse data. The rapid PA 
change and modal connecting bridge are not apparent in the 430-MHz polarization 
observations reported by BRC, as one might expect from the frequency 
independence of the RC model. The PAs of PSR B2016+28 observed at 1404 MHz 
then pose a multi-faceted dilemma. If the RC model is to be preserved, one 
must decide which viewing geometry is correct while also developing an 
explanation for the PA trace that is not attributed to geometry. Furthermore, 
one must determine the origin of the modal connecting bridge. Is it an 
artifact of polarization mode interaction or could it be an indicator of 
another emission mechanism operative in the pulsar magnetosphere?

  Despite the conflicting interpretations generated by the 1404-MHz PAs of PSR 
B2016+28, its viewing geometry has generally been accepted on the basis of 
the frequency dependence of its pulse shape. Rankin (1983, 1986, 1993) developed 
a morphological classification scheme that is based primarily on the spectral 
evolution of pulse shapes. She categorizes PSR B2016+28 within a group of pulsars 
designated as \lq\lq conal single" or \lq\lq type-Sd" pulsars. The pulse shape of 
a conal single pulsar evolves from a single component at high frequency to two 
components at low frequency. This frequency dependence of the pulse shape is 
thought to result from a combination of the spreading or radius-to-frequency 
mapping of the conal emission and a tangential traverse of an observer's 
sightline across the pulsar's emission cone. The viewing geometry for conal 
single pulsars thus requires a large angle between the magnetic pole and the 
sightline, just as the modal PA trajectories in PSR B2016+28 suggest. The pulsar 
displays other key characteristics of conal single pulsars (Rankin 1983; Rankin 
\& Ramachandran 2003), such as drifting subpulses and low linear polarization 
across its entire pulse owing to orthogonally polarized modes (OPM). Hankins 
\& Rickett (1986) investigated the frequency dependence of the average profiles 
for many pulsars. Specifically recognizing the conflicting interpretations of 
the 1404-MHz PAs in PSR B2016+28, they noted that the modal PA trajectories in 
the pulsar are more consistent with the frequency dependence of its profile. 
Hankins et al. (1992) interpret the 1404-MHz PAs, and thus the pulsar's viewing 
geometry, in a similar way. However, no one has adequately resolved the dilemma 
over the conflicting interpretations of the 1404-MHz PAs or reconciled, in 
detail, the 1404-MHz PAs with the viewing geometry implied by its morphological 
classification. Additionally, no one has offered a detailed explanation for the 
pulsar's modal connecting bridge.

  SCRWB suggested that the modal connecting bridge in PSR B2016+28 and, more
generally, the nonorthogonality of the polarization modes may be related to the 
\lq\lq migration" of OPM. This mode migration may be the frequency-dependent 
relocation, spreading, or splitting of modal radiation with pulse longitude. 
SCRWB also emphasized the need to clarify the nature and importance of 
nonorthogonal polarization modes (NPM) in the emission. While the occurrence of 
NPM has been frequently observed and widely documented in the literature (e.g. 
BR; SCRWB), theoretical investigations (e.g. Cheng \& Ruderman 1979; Allen \& 
Melrose 1982; Barnard \& Arons 1986; Michel 1987; Gangadhara 1997) and 
statistical modeling (e.g. McKinnon \& Stinebring 1998, 2000; McKinnon 2002) of 
the modes have concentrated on their orthogonality, but have largely ignored 
their observed nonorthogonality.

  The first objective of this paper is to resolve the dilemma over the
conflicting interpretations of the pulsar's 1404-MHz PAs. I follow the suggestion 
of SCRWB and argue that the rapid change in average PA along with the modal 
connecting bridge in PSR B2016+28 is a longitudinally-resolved transition between 
NPM. In other words, the bridge is simply an artifact of NPM superposition, and 
not an indicator of another emission mechanism. The second objective of the paper 
is to account for nonorthogonal polarization modes in a statistical model of the 
radio emission's polarization. In \S\ref{sec:obs}, I review the observed 
polarization properties of PSR B2016+28 by replicating the results of SCRWB, and 
argue that the modal PA trajectories reflect the true viewing geometry of the 
pulsar. I incorporate the mode nonorthogonality in a statistical model of 
superposed OPM (McKinnon \& Stinebring 1998, hereafter MS1) in \S\ref{sec:nonortho}. 
I then describe the predicted effects of superposed NPM on polarization data and 
compare the predictions with observations. In \S\ref{sec:discuss}, I discuss the 
implications of this investigation for radiative processes in pulsar magnetospheres. 

\section{OBSERVED POSITION ANGLES IN PSR B2016+28}
\label{sec:obs}

\begin{figure}
\plotone{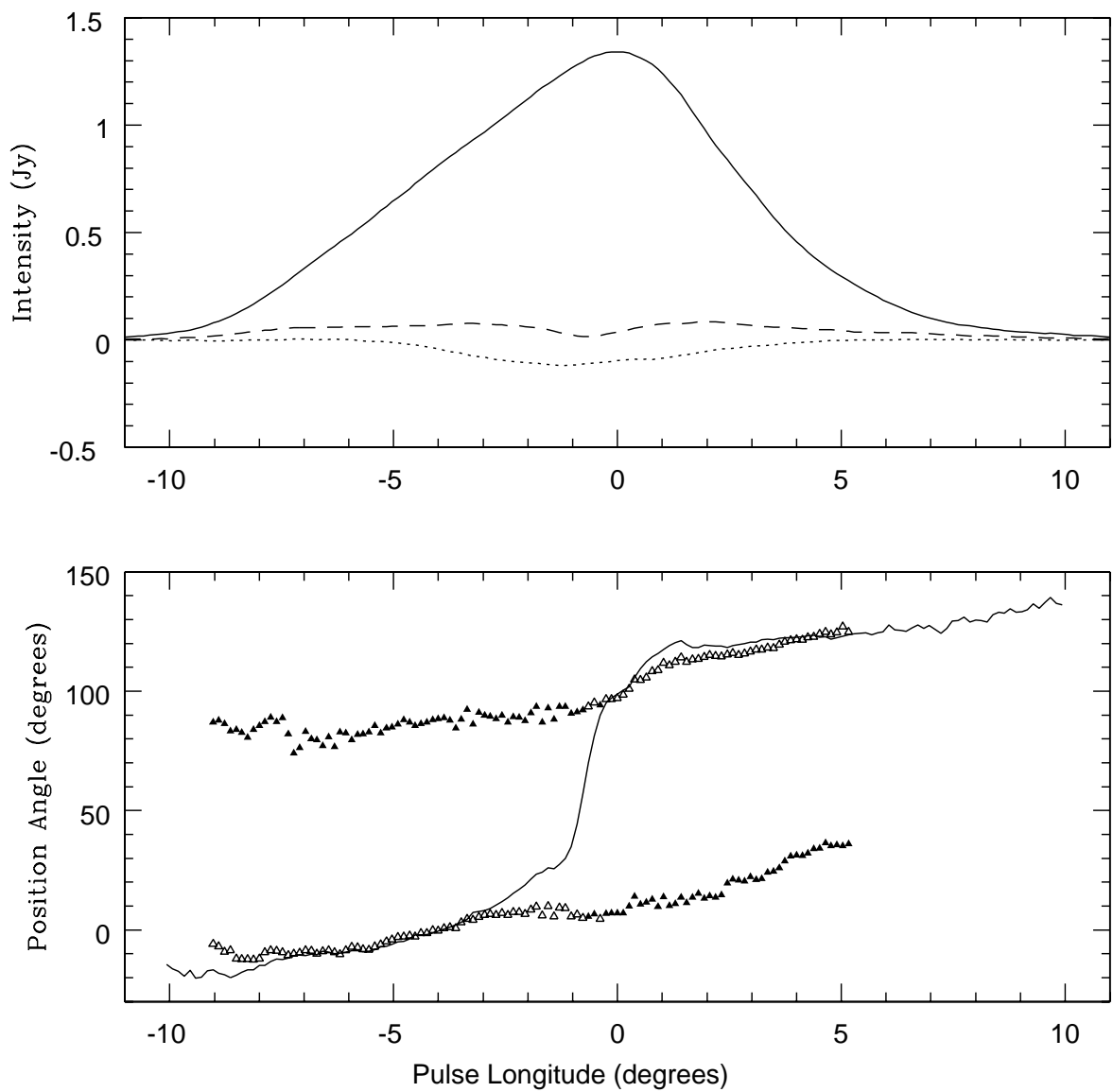}
\caption{Average pulse profile of PSR B2016+28 at 1404 MHz. The total 
intensity, linear polarization, and circular polarization of the pulse are 
denoted by the solid, dashed, and dotted lines, respectively, in the top
panel. The polarization position angle (PA) determined from the average values 
of the Stokes parameters Q and U is shown by the solid line in the bottom 
panel. The triangles in the PA plot denote the peaks in PA histograms constructed 
from single pulse polarization observations. Open triangles represent the 
dominant polarization mode, and filled triangles represent the weak mode.}
\label{fig:profile}
\end{figure}

  As a point of reference for the discussion in \S~\ref{sec:nonortho}, the 
observational results of SCRWB are replicated in Figure~\ref{fig:profile}, which 
shows the average pulse profile of PSR B2016+28 at 1404 MHz. The single pulse, 
polarization data used to construct the profile were recorded by SCRWB with the 
305-m Arecibo radio telescope. The solid line in the bottom panel of the figure 
shows how the average PA varies across the pulse. The PAs of the pulsar's 
nearly-orthogonal polarization modes at each pulse longitude are shown by the 
triangles in the PA plot, and were estimated as follows. 

  A PA histogram, similar to those shown in Figure~\ref{fig:paquad}, was 
computed at each pulse longitude of PSR B2016+28 using the single pulse, 
polarization observations of SCRWB. Each histogram contained 50 
equally-spaced bins over the $180^\circ$ range in possible PA values, giving a 
histogram bin resolution of $3.6^\circ$. Only data samples having a linear 
polarization that exceeded the off-pulse noise by a factor of five were used to 
construct the histograms. The PA of the most frequently occurring, or dominant, 
polarization mode was found by locating the peak in the histogram, and using the 
peak bin along with 12 bins on either side of it to calculate a weighted mean 
angle, $\langle\psi\rangle$. From the statistics of directional data described 
in Mardia (1972), the weighted mean angle was found using

\begin{equation}
C={1\over {N}}\sum_{i=1}^{25} f_i\cos(2\psi_i),
\label{eqn:C}
\end{equation}

\begin{equation}
S={1\over {N}}\sum_{i=1}^{25} f_i\sin(2\psi_i),
\label{eqn:S}
\end{equation}

\begin{equation}
\langle\psi\rangle = {1\over {2}}\arctan\Biggl({S\over {C}}\Biggr),
\label{eqn:meanpsi}
\end{equation}
where $\psi_i$ is the midpoint of the PA bin, $f_i$ is the number of data 
samples in the bin, and $N=\sum f_i$ is the total number of data samples in the 
25 PA bins. The open triangles in Figure~\ref{fig:profile} show the weighted 
mean angles for the dominant mode at each pulse longitude. The remaining 25 bins 
in the PA histogram were used to calculate the mean angle of the weak polarization 
mode (filled triangles in the figure) with the same equations. The resulting 
PA trajectories of the polarization modes are very similar to those reported in 
the original observations (see Fig. 31 of SCRWB).

  Of the 2700 pulses recorded in the observation, the number of data samples
exceeding the detection threshold on linear polarization was as high as 1250, 
depending on pulse longitude. Data points are not shown in the bottom panel 
of Figure~\ref{fig:profile} for any pulse longitude having less than 100 
samples in its PA histogram. These locations occur at the pulse edges where the 
instantaneous signal-to-noise ratio (S/N) in linear polarization is low. Formal 
errors in $\langle\psi\rangle$ are estimated to range from $6^\circ$ to 
$13^\circ$, and have been omitted from the figure for clarity.

  The primary advantage of calculating modal PAs from equations~\ref{eqn:C}, 
\ref{eqn:S}, and~\ref{eqn:meanpsi} is to bring to bear the full weight of the 
data in making quantitative estimates of modal PAs and their differences, thus 
helping to identify any deviations from mode orthogonality. A minor disadvantage 
of this particular method is to introduce a slight bias that skews the difference 
in modal PAs towards $90^\circ$ when the PA is nearly uniformly distributed and 
the linear polarization is very low (e.g. on the pulse edges and near a pulse 
longitude of $-1^\circ$ in Fig.~\ref{fig:profile}).

  The PAs in the bottom panel of Figure~\ref{fig:profile} clearly illustrate 
the two interpretations that could be made of the pulsar's viewing geometry. The 
average angle changes rapidly with pulse longitude near the pulse center, and its 
total excursion across the pulse approaches $180^\circ$. Both properties are 
indicative of a viewing geometry where the observer's sightline passes near the 
star's magnetic pole. The trajectories followed by the polarization modes, 
however, are relatively flat with a total excursion in PA of something much less 
than $180^\circ$, indicating that the observer's sightline is far from the 
magnetic pole. 

  Many features of Figure~\ref{fig:profile} suggest that the flat PA
trajectories of the polarization modes reflect the pulsar's true viewing 
geometry. The average PA generally follows the dominant polarization mode (open 
triangles in Fig.~\ref{fig:profile}) as one would expect from OPM superimposed 
upon a flat PA trajectory (BRC; SCRWB; MS1). One polarization mode dominates the 
leading edge of the pulse, while the other dominates the trailing edge. Assuming 
that the modes occur simultaneously (MS1; McKinnon \& Stinebring 2000, hereafter 
MS2), the low average linear polarization and the frequent occurrence of both 
modes across the entire pulse imply that the modes have comparable polarized 
intensities on average (see, also, Rankin 1983; Rankin \& Ramachandran 2003). 
Additionally, the rapid change in average angle near the pulse center isn't quite 
as smooth as one would typically expect from the RC model. The rapid PA change 
joins the modal PA trajectories, and its magnitude is approximately $90^\circ$, 
suggesting that the PA change is related to an OPM transition. 

  Additional features of Figure~\ref{fig:profile} suggest that NPM may be 
complicating our interpretation of the PA pattern in PSR B2016+28. The 
differences between the modal PAs are almost, but not precisely, $90^\circ$ 
across most of the pulse. The mode nonorthogonality does not appear to vary 
greatly or rapidly across the pulse. While the change in the average angle at 
the pulse center is rapid, it is not discontinuous as would be expected for a 
purely orthogonal transition, and so may indicate the presence of NPM (Cheng 
\& Ruderman 1979; SCRWB; Michel 1987).

\section{STATISTICAL MODEL OF NONORTHOGONAL MODES}
\label{sec:nonortho}

  A statistical model for the polarization of pulsar radio emission was presented
in MS1 and further developed in MS2 and McKinnon (2002). The model is based upon 
the observation that the highly polarized, orthogonal modes occur simultaneously. 
To incorporate the nonorthogonality of the modes in the model, the analysis which 
follows concentrates on the emission's linear polarization and neglects its 
circular polarization because NPM are most evident in the observed PA of the 
linear polarization vector and the circular polarization generally tends to be a 
small part of the emission's polarization. 

\begin{figure}
\plotone{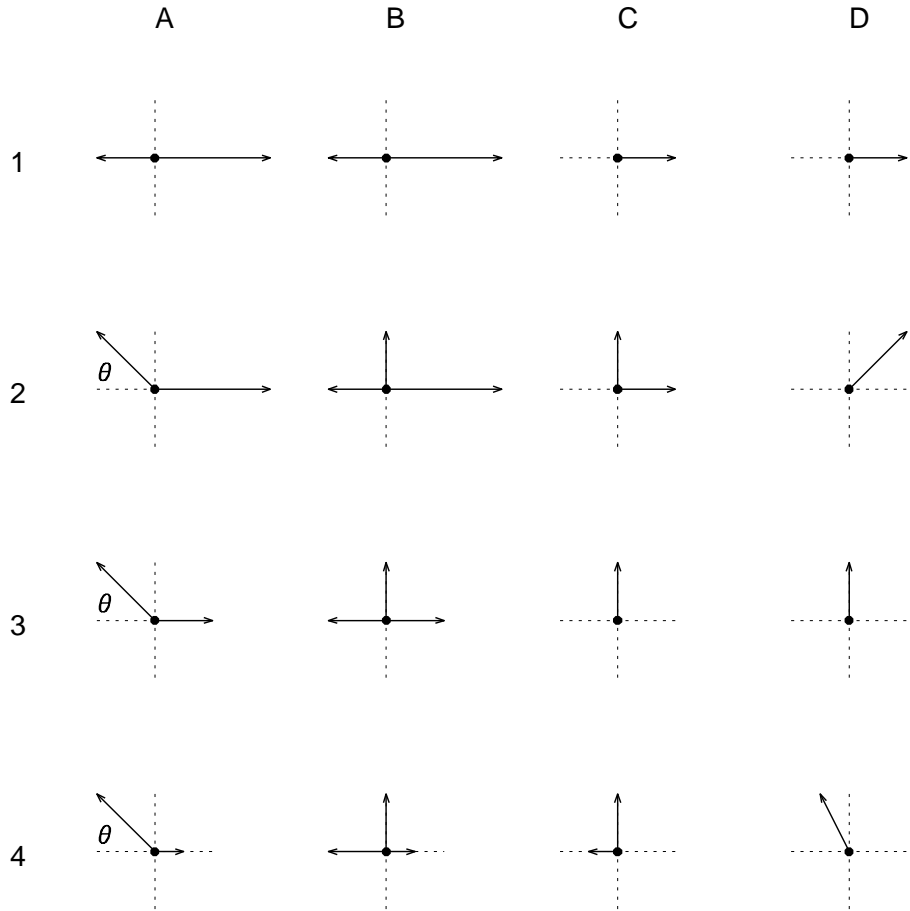}
\caption{Four examples (1-4) of the polarization vector resulting from 
the superposition of two polarization modes. In each example, column A 
shows the vectors representing the linear polarizations of the primary 
and secondary modes. The primary mode vector points to the right in each 
grid of column A, and the secondary mode vector points towards the left, 
generally making an angle $\theta$ with the horizontal. Column B shows 
the primary mode and the decomposition of the secondary mode into its 
orthogonal components. Column C shows the vector sums of the orthogonal 
components. The resultant vector is shown in column D. The modes are 
orthogonal in example 1. Examples 2 through 4 illustrate the effect of 
decreasing the amplitude of the primary mode when the modes are not 
orthogonal.}
\label{fig:vectors}
\end{figure}

  As in MS1, let $X_1$ and $X_2$ be random variables representing the linearly 
polarized intensities of the primary and secondary polarization modes, 
respectively. Without loss of generality, the PA of the primary 
mode can be set to $0^\circ$, which means that the polarization vector of the 
primary mode points in the direction of increasing Stokes parameter Q in the 
Q-U plane of the Poincar\'e sphere (Fig.~\ref{fig:vectors} ). The polarization 
vector of the secondary mode generally points in the direction opposite that of 
the primary mode vector, but makes an angle $\theta$ with respect to the Q-axis 
to account for the nonorthogonality (see Fig.~\ref{fig:vectors}). If 
$X_{\rm N}$ is a random variable representing the instrumental noise, the 
observed Stokes parameters Q and U of the combined emission at an arbitrary 
pulse longitude are
\begin{equation}
{\rm Q} = X_1 - X_2\cos\theta + X_{\rm N,Q},
\label{eqn:Q_model}
\end{equation}
\begin{equation}
{\rm U} = X_2\sin\theta + X_{\rm N, U}.
\label{eqn:U_model}
\end{equation}
Assuming that $\theta$ is constant and thus independent of $X_2$, the PA 
calculated from the average values of Q and U is
\begin{equation}
\psi= {1\over{2}}\arctan\Biggl({\langle {\rm U}\rangle\over 
      {\langle {\rm Q}\rangle}}\Biggr) =
{1\over{2}}\arctan\Biggl({\sin\theta\over {M-\cos\theta}}\Biggr),
\label{eqn:psi}
\end{equation}
where $M=\mu_1/\mu_2$ is the ratio of the mean values of the modes' polarized 
intensities, $\mu_1=\langle X_1\rangle$ and $\mu_2=\langle X_2\rangle$. 
Equation~\ref{eqn:psi} shows that the average PA resulting from the superposition 
of NPM produces an angle intermediate to the PAs of the two modes. For example, 
when $M=1$ and the nonorthogonal modes have PAs of $0^\circ$ and $85^\circ$, the 
deviation from orthogonality is $\theta=10^\circ$ and the average angle is 
$\psi=42.5^\circ$, which is equidistant from the orientations of the two modes.

  If the modes are truly orthogonal ($\theta = 0$), the vectors representing 
their linear polarization are antiparallel in the Q-U plane. The polarization 
vector resulting from the superposition of OPM is aligned with the PA
of the mode having the larger polarized intensity (Fig.~\ref{fig:vectors}, 
example 1). The MS1 statistical model attributes the observed switching between 
modes to the random fluctuations in their polarized intensities. Within the 
context of the model, the only way to produce PAs intermediate to those of the 
orthogonal modes is through the influence of the instrumental noise. PA 
histograms produced by this model generally contain two peaks, symmetric about 
their centroids and separated by $90^\circ$, with a uniform plateau arising 
from the instrumental noise (see Fig. 2 of MS1). 

  When the modes are not orthogonal, their polarization vectors are not
antiparallel in the Q-U plane. As can be seen in examples 2 through 4 of 
Figure~\ref{fig:vectors}, the orientation of the vector resulting from NPM
superposition is intermediate to the orientations of the mode vectors, and 
depends upon the modes' polarized intensities and the degree of nonorthogonality 
($\theta$ in Fig.~\ref{fig:vectors}). Also, the superposition of NPM depolarizes 
the observed emission (i.e. the amplitude of the resultant polarization vector 
is less than that of the polarization vector for the dominant mode), but not 
nearly as efficiently as the superposition of OPM. 

\begin{figure}
\plotone{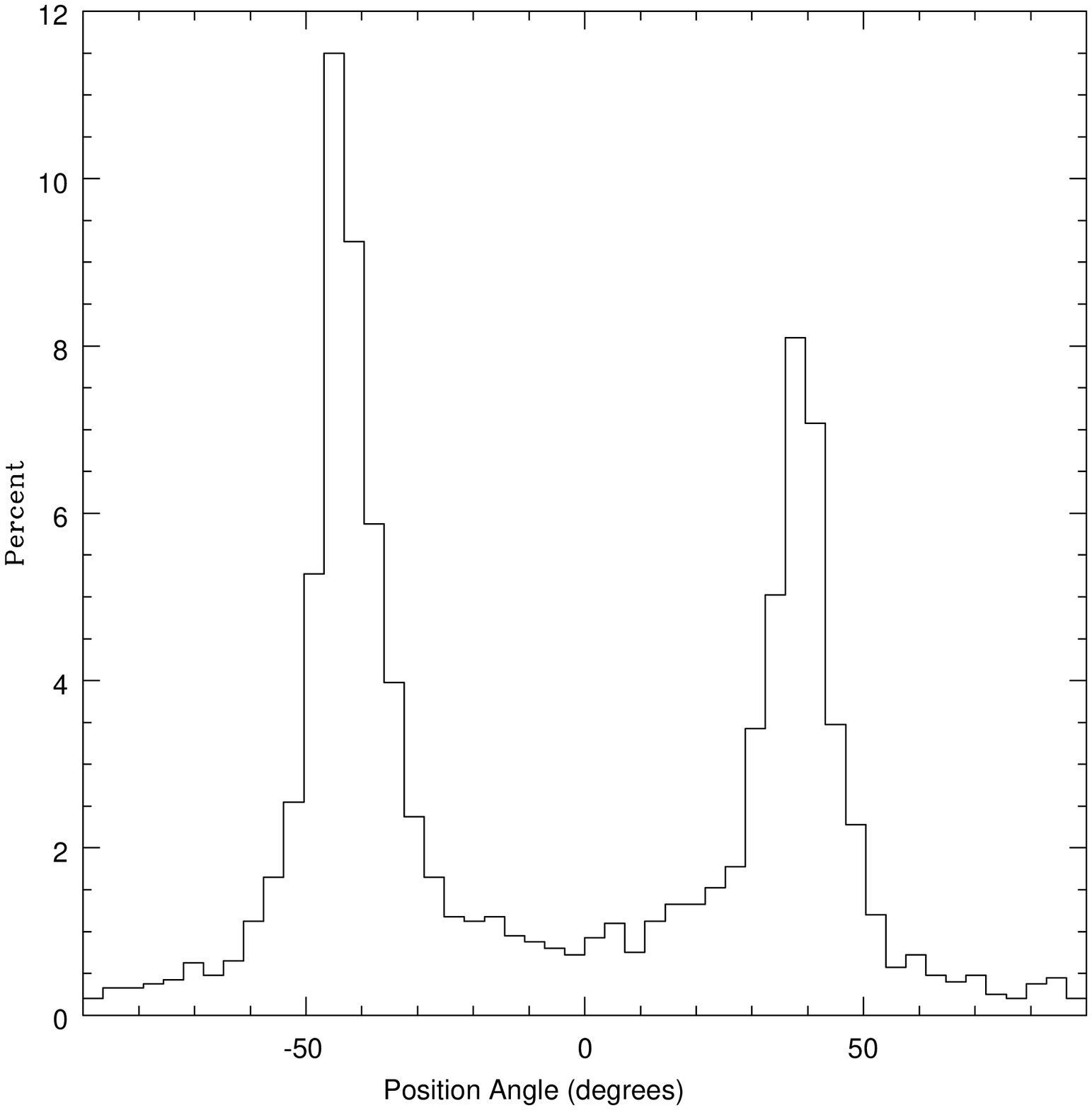}
\caption{Position angle histogram constructed from a numerical simulation
of superposed, nonorthogonal modes. The histogram illustrates the
characteristics expected from the superposition of nonorthogonal 
modes (see the text). The mode polarizations were modeled as exponential 
random variables, and instrumental noise was included in the simulation.
The degree of nonorthogonality for the secondary mode was $\theta=10^\circ$ 
in the simulation. The peak of the primary polarization mode has been set 
to $-45^\circ$ for display purposes.}
\label{fig:PAmodel}
\end{figure}

  These signatures of the superposed NPM model should be evident in PA histograms
and in the joint probability density, or scatter plots, of PA and linear 
polarization (L). The PA histograms should be similar to Figure~\ref{fig:PAmodel}, 
which was constructed from a numerical simulation of equations~\ref{eqn:Q_model} 
and~\ref{eqn:U_model}. The separation of the peaks in the histogram is less than 
$90^\circ$ because the modes are not orthogonal. Since PAs intermediate to the 
orientations of the mode vectors can result from the superposition of NPM, an 
overabundance of data samples occurs between the peaks, and a paucity of data 
points occurs outside them. The intermediate values of the resultant PAs also 
lead to slight asymmetries in the peaks about their centroids. For the L-PA 
scatter plots, the NPM model predicts that data samples with significant L should 
occur between the mode peaks.

\begin{figure}
\plotone{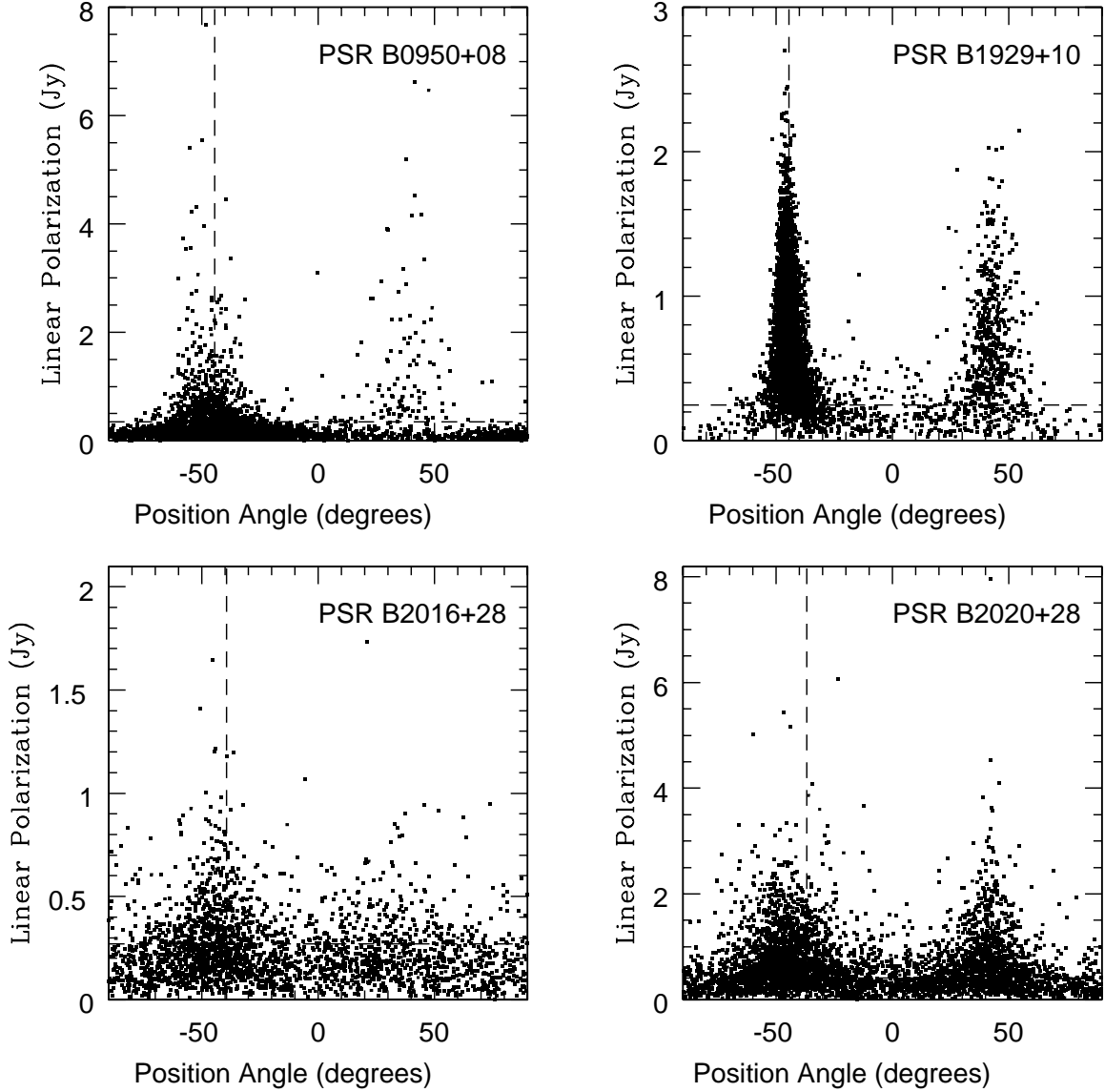}
\caption{Scatter plots of position angle (PA) and linear polarization (L) at 
a single pulse longitude in each of PSR B0950+08, PSR B1929+10, PSR B2016+28, 
and PSR B2020+28. The dashed horizontal line in each plot is drawn at a power 
level corresponding to five times the off-pulse noise. Many data samples with 
significant L, particularly in PSR B0950+08 and PSR B1929+10, are located 
between the mode peaks as predicted by the NPM statistical model. The vertical 
line denotes the location of the PA computed from the average values of Stokes
Q and U. For display purposes, the original values of Q and U were rotated so 
that the PA of the dominant polarization mode occurs at $-45^\circ$.}
\label{fig:joint}
\end{figure}

  Figure~\ref{fig:joint} shows L-PA scatter plots for four pulsars observed 
by SCRWB. In each plot, the dominant polarization mode is located where the 
concentration of data points is greatest. The observed Stokes parameters Q 
and U of each data point have been rotated so that the dominant mode occurs 
at a PA of $\psi = -45^\circ$ for each pulsar. As predicted by the superposed 
NPM model, many data samples with highly significant linear polarization 
(${\rm L} > 5\sigma_N$) are located between the mode peaks 
($-45^\circ < \psi < 45^\circ$). This is particularly true of PSR B0950+08 and 
PSR B1929+10 where the modes are clearly defined. The modes are not as well 
defined in the scatter plots for PSR B2016+28 and PSR B2020+28 because the modes
occur with similar frequency. As discussed in MS1, this implies the mode 
polarization amplitudes are similar on average (i.e. $M\simeq 1$), so that their 
simultaneous interaction leads to significant depolarization of the intrinsic 
radiation and causes the observed polarization to be dominated by the randomly 
polarized, on-pulse, instrumental noise (the broad pedestal in the scatter plots). 

\begin{figure}
\plotone{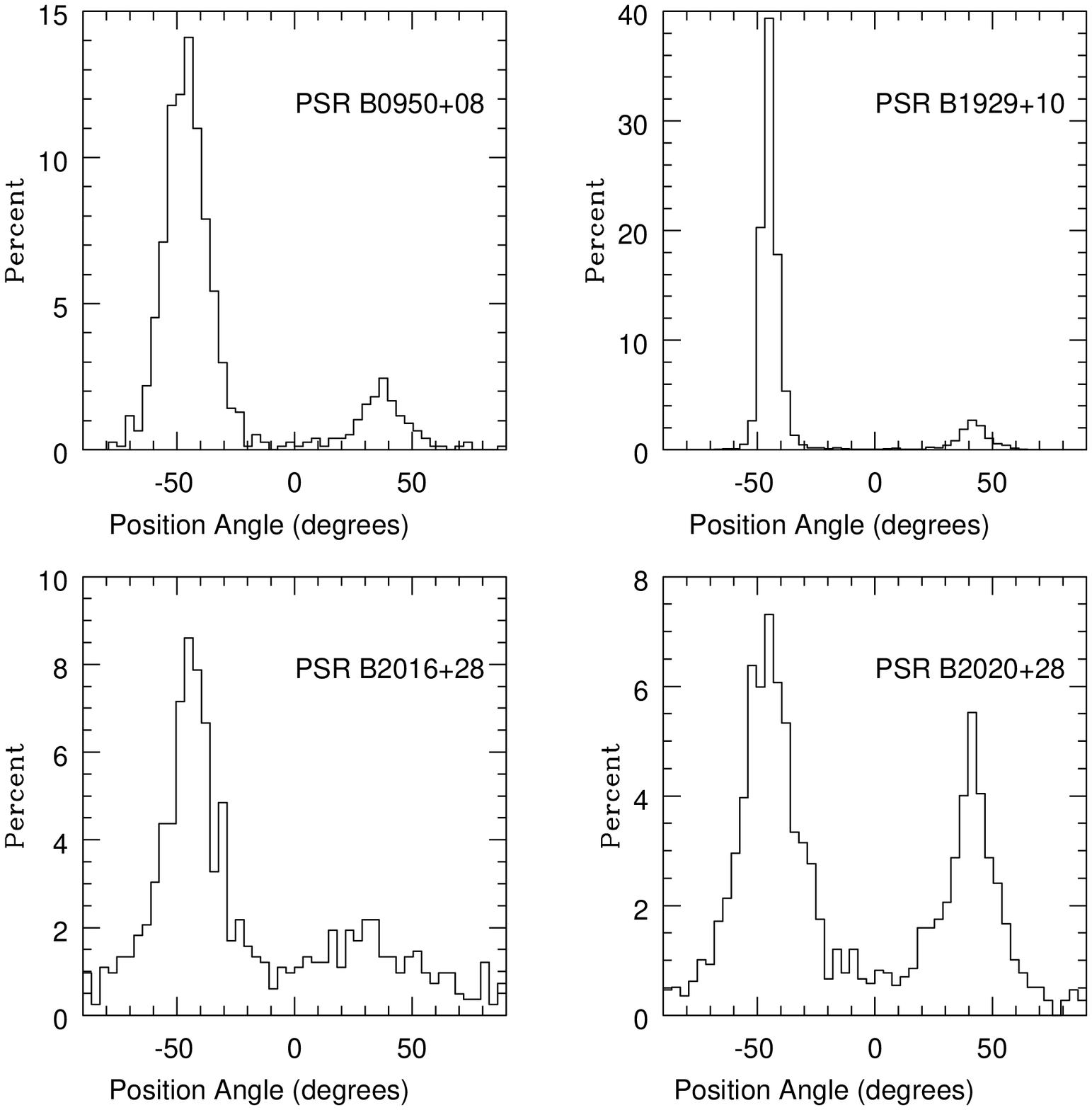}
\caption{Position angle (PA) histograms constructed from the scatter plots in 
Fig.~\ref{fig:joint}. In each case, the separation in histogram peaks is less 
than $90^\circ$, indicating that the polarization modes in each pulsar are not 
orthogonal. The peak of each histogram, which occurs at the PA of the dominant 
polarization mode, has been centered on $-45^\circ$ for display purposes.}
\label{fig:paquad}
\end{figure}

  Figure~\ref{fig:paquad} shows the PA histograms constructed from the scatter 
plots in Figure~\ref{fig:joint}. As mentioned in \S~\ref{sec:obs}, only data 
points with a S/N in L greater than five (the dashed horizontal line in each 
plot of Fig.~\ref{fig:joint}) were used to construct the histograms. The 
separation between the two peaks in each histogram is less than $90^\circ$, and 
of the four examples shown in the figure, the deviation from mode orthogonality 
is greatest for PSR B2016+28. The four histograms show other similarities in 
addition to the nonorthogonality of the modes. In each histogram, more data points 
fall between the peaks than outside them, and the individual peaks are slightly 
asymmetric about their centroids. The PA histograms bear a remarkable resemblance 
to the histogram generated with the numerical simulation of the superposed NPM 
model (Fig.~\ref{fig:PAmodel}).

  Turning now to the PA calculated from the averaged Stokes parameters, 
consider a hypothetical situation where one had {\it apriori} knowledge of 
a pulsar's viewing geometry. This being the case and barring the limitations of 
S/N, one could easily remove the systematic effect of the RC PA sweep from the 
data. Having done this, the variation in PA across the pulse attributable to the 
polarization modes would be described by equation~\ref{eqn:psi}, where $M$ 
and $\theta$ are now dependent on pulse longitude. The primary mode would 
occur at pulse longitudes having $M>\cos\theta$, the secondary mode would 
occur at longitudes with $M<\cos\theta$, and a mode transition would occur 
at longitudes where $M=\cos\theta$. If the deviation from orthogonality is 
small and constant ($\theta=\theta_o << 1$) over the range of pulse longitude 
where the mode transition occurs and if the modal ratio, $M$, varies linearly 
with longitude (e.g. $M(\phi) = 1 + a(\phi - \phi_o)$) over the same range, 
the rate the PA changes with pulse longitude near the mode transition is

\begin{equation}
{d\psi\over {d\phi}} \simeq -{a\over{2\theta_o}}.
\label{eqn:slope}
\end{equation}

\noindent Equation~\ref{eqn:slope} shows that the mode transition is abrupt 
when the modes are orthogonal ($\theta_o=0$), but gradual when the modes are 
not orthogonal ($\theta_o\ne 0$). The equation also suggests that a mode 
transition may be longitudinally resolved provided that $\theta$ and $M$ vary 
slowly with pulse longitude, which is likely the case for PSR B2016+28 for 
reasons cited in \S~\ref{sec:obs}. 

\begin{figure}
\plotone{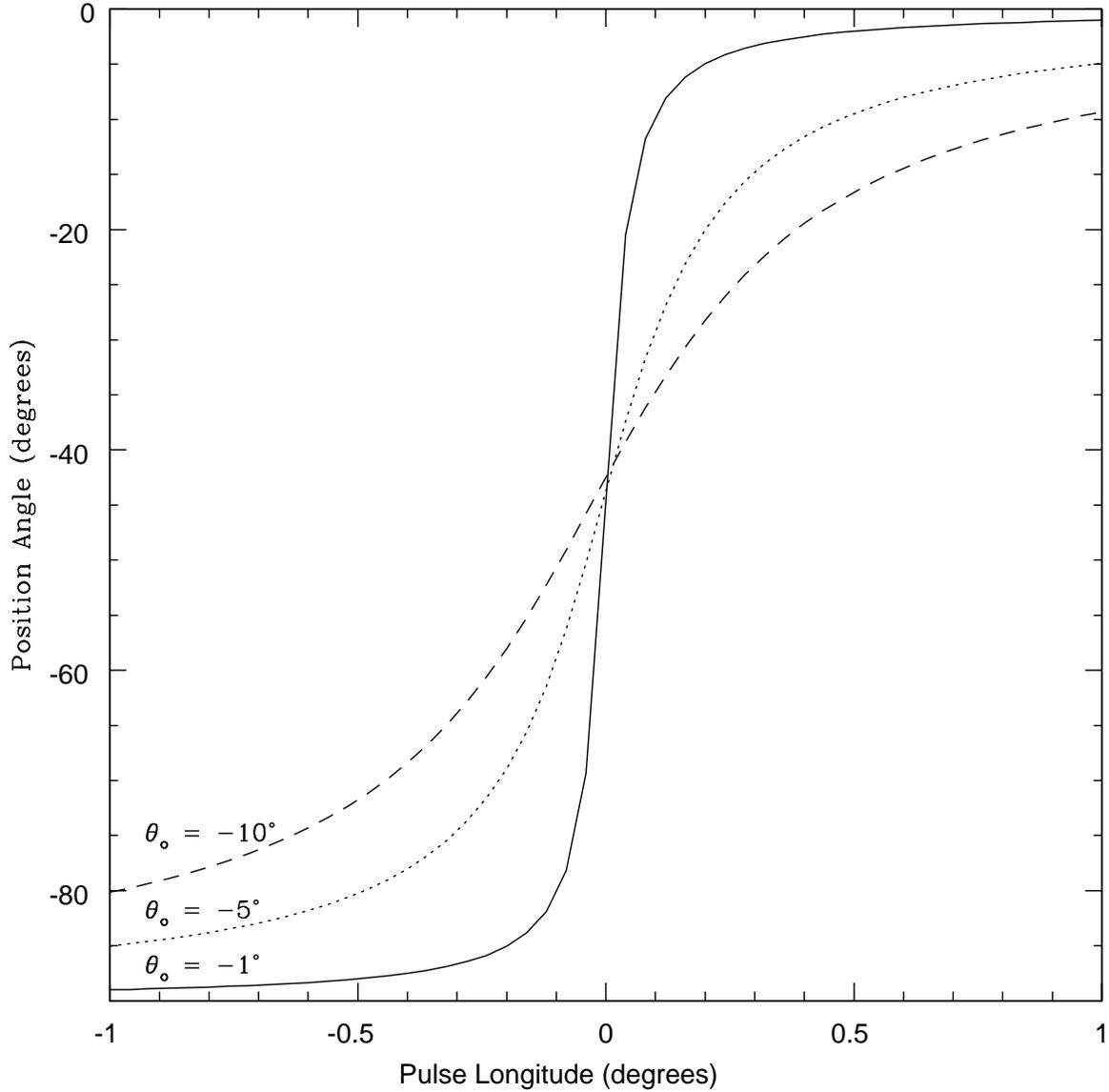}
\caption{Three examples of longitudinally-resolved position angle transitions 
due to the superposition of nonorthogonal polarization modes. The transitions
are more abrupt when the deviation from mode orthogonality, $\theta_o$, is
small.}
\label{fig:moderatio}
\end{figure}

  The PA change with pulse longitude that is described by equation~\ref{eqn:psi} 
is shown in Figure~\ref{fig:moderatio} for different values of $\theta_o$, with 
$M$ varying linearly with longitude. The change in average PA predicted by the 
statistical model (Fig.~\ref{fig:moderatio}) is qualitatively consistent with 
what is observed in PSR B2016+28 (cf. bottom panel of Fig.~\ref{fig:profile}). 
Equation~\ref{eqn:psi} and Figure~\ref{fig:moderatio} also illustrate the point 
that a PA transition between purely orthogonal modes will be discontinuous and 
instantaneous, whereas nonorthogonal modes will produce a smooth or gradual PA 
transition (Cheng \& Ruderman 1979; SCRWB; Michel 1987).

  If, as in SCRWB, one were to convert the PA histograms at each pulse longitude 
(e.g. Fig.~\ref{fig:PAmodel}) to grey-scale histograms and overlay them on the 
averaged PA data (e.g. Fig.~\ref{fig:moderatio}), the average angle would be 
embedded in the pedestal that appears between the peaks in each histogram because 
the PA resulting from the superposition of NPM is intermediate to the mode PAs 
(Eq.~\ref{eqn:psi}). In overlaying all the histograms on the average data, the 
individual pedestals combine to form a much broader pedestal that spans a large 
range in pulse longitude. This broad pedestal is the modal connecting bridge in 
PSR B2016+28 discovered by SCRWB (see Fig. 31 of SCRWB).

  The preceding point is also illustrated in Figure~\ref{fig:joint} where actual 
average angles are shown by the dashed vertical lines in each L-PA scatter plot. 
In each case, the average angle is not precisely aligned with the dominant mode 
PA ($\psi = -45^\circ$). The difference between the dominant mode PA and the
average PA, $\Delta\psi$, is greatest for PSR B2016+28 and PSR B2020+28 where the 
modes occur with similar frequency and the mode nonorthogonality is significant. 
This is to be expected from the model because equation~\ref{eqn:psi} predicts 
that $\Delta\psi$ is large for $M\simeq 1$ and $\theta > 0$.

\section{DISCUSSION}
\label{sec:discuss}

  The statistical model's ability to simulate the rich variety of polarization 
properties observed in pulsar radio emission provides strong support for the 
model's underlying assumption of superposed polarization modes. Observations of 
PSR B2020+28 support the model's prediction that the emission will be depolarized 
on all timescales and that the modes will occur with nearly equal frequency when 
the polarized intensities of the orthogonal modes are comparable (MS1; MS2). 
Additionally, the model adequately reproduces observed histograms of total 
intensity, linear polarization, fractional linear polarization, and PA in PSR 
B2020+28 (MS1; MS2). Since the polarization modes are superposed and elliptically 
polarized in general, the model also predicts that histograms of fractional 
circular polarization should be unimodal, in accord with the observations 
(McKinnon 2002). If the modes occurred separately, the histograms of fractional 
circular polarization would be bimodal, which is not observed. In this paper, the 
model resolves the longstanding dilemma over the interpretation of the 1404-MHz
PAs in PSR B2016+28, accounts for its modal connecting bridge, and explains many 
subtle features of PA histograms by incorporating the nonorthogonality of the 
modes.

  Obvious questions that arise from an investigation of NPM are: \lq\lq what 
is their origin" and \lq\lq how does the statistical model comport with the 
explanation of NPM generation"? The task of explaining the origin of OPM is 
daunting enough by itself, and little attention has been devoted to NPM's 
origin because it has been regarded as a minor or subtle aspect of OPM 
generation. OPM may be intrinsic to the emission mechanism (e.g. Gangadhara 
1997; Cheng \& Ruderman 1979) or produced at widely separated regions which 
happen to have magnetic field structure that is orthogonal when projected on 
the plane of the sky (MTH; Michel 1987). However, the most widely accepted 
explanation for OPM appeals to propagation effects in the magnetized plasma 
above the pulsar polar cap (e.g. Allen \& Melrose 1982; Barnard \& Arons 1986). 
The radio emission can be spatially separated into two modes of orthogonal 
polarization due to the difference in their indices of refraction. If the 
refraction is strong, the emission in one mode that is generated at one 
location within the magnetosphere may overlap the emission from the other 
mode that is generated at a different location. In this particular scenario 
which is entirely consistent with SCRWB's suggestion that NPM arise from 
OPM migration, the polarizations of the overlapping modes are not orthogonal 
because the modes were not generated at the same location. The observed 
polarization resulting from the overlap of the two nonorthogonal modes is 
precisely what is illustrated in Figure~\ref{fig:vectors} and described by 
equations~\ref{eqn:Q_model} and~\ref{eqn:U_model}. Therefore, the statistical 
model is consistent with the proposition that OPM and NPM arise from 
propagation effects in the pulsar magnetosphere. More generally, the model 
is consistent with any mode production mechanism that involves the incoherent 
superposition of the modes. The exact details of how OPM migration actually 
occurs in PSR B2016+28 are unknown, but may be related to the pulsar's 
drifting subpulses (Drake \& Craft 1968; Backer 1973) and the occurrence 
of OPM within them, as in other conal single pulsars such as PSR B0943+10 
(Deshpande \& Rankin 2001) and PSR B0809+74 (Ramachandran et al. 2002).

  In an admirable first attempt to explain NPM, SCRWB modeled the emission's
polarization as the sum of a nonmodal component and OPM. While SCRWB's model 
cannot be ruled out, it introduces an additional level of complexity for 
radiative processes in pulsar magnetospheres by requiring the generation
of both a nonmodal emission component and OPM. The statistical model, by
contrast, portrays a much simpler situation where radiative processes are
only required to produce two polarization modes, either at emission or through 
propagation, with slightly nonorthogonal polarizations. Furthermore, the model 
leads to a simple explanation for the modal connecting bridge in PSR B2016+28. 
Any other explanation for the bridge would likely resort to rather {\it ad hoc} 
assumptions about its origin, placing additional demands on theoretical 
explanations of radiative processes in the magnetosphere as a result. 

  A number of observations have posed serious challenges to the validity of the 
RC model, but the model has generally withstood the test of the observations.
The polarization observations of the drifting subpulses in PSR B0809+74 by 
Taylor et al. (1971) indicated that PA variations were synchronized with the 
subpulses, and not fixed to the rotation of the star as the RC model predicted. 
Only recently have Ramachandran et al. (2002) shown that the PA variations within 
the subpulses are due to OPM transitions, and that the individual modes follow 
PA trajectories that are consistent with the RC model. Initial observations of 
OPM (e.g. MTH) also challenged the RC model because it could not account for the 
production of the modes. However, BRC noted from their single pulse, polarization 
observations of PSR B2016+28 that the RC model was relevant because the modal PA 
traces followed parallel trajectories that were consistent with the model. SCRWB 
pointed out that the 1404-MHz PAs of PSR B2016+28 could be interpreted in the 
context of the RC model with two, very different viewing geometries. In this 
paper, the PA interpretation dilemma is resolved and the RC model is thus 
preserved by accounting for the nonorthogonality of the pulsar's polarization 
modes.

  The occurrence of NPM has important implications for the interpretation of
pulsar polarization and viewing geometry. As shown in SCRWB and this paper, 
the occurrence of NPM in PSR B2016+28 so disrupts the PA sweep that two, 
very different interpretations can be made of its viewing geometry. Everett 
\& Weisberg (2001) used some of the most sensitive polarization observations 
recorded to date in an attempt to tightly constrain the viewing geometry of a 
number of nearby pulsars. Interestingly, the best fits of their data to the RC 
model were obtained by completely discarding data recorded on the pulse and 
relying on data of lower S/N recorded off the pulse. The viewing geometries 
they determined for PSR B0950+08 and PSR B1929+10 were very different from 
results obtained in previous work. The fact that both pulsars exhibit NPM 
(Fig.~\ref{fig:paquad}; BR; SCRWB) may explain why data recorded on their pulses 
deviated significantly from the RC model. Deviations from the RC model have 
traditionally been attributed to multi-polar magnetic field structure,
implying that the emission region is close to the stellar surface. But if 
the deviations from the RC model are caused by NPM, as suggested here, the 
emission may originate at a higher location in the magnetosphere where the 
star's dipole component dominates the multi-pole components of its magnetic 
field. Considering the substantial impact of NPM on the PA sweep in PSR 
B2016+28, propagation effects may also play a more significant role in 
determining the observed polarization than other mechanisms, such as 
relativistic aberration (Blaskiewicz, Cordes, \& Wasserman 1991) and current 
flow above the polar cap (Hibschman \& Arons 2001).

  In summary, I have incorporated the nonorthogonality of the polarization modes 
in a statistical model of the polarization of pulsar radio emission. I used the 
model to show that the rapid change in average PA near the pulse center of PSR 
B2016+28 at 1404 MHz may be a longitudinally-resolved transition betweeen modes 
of nonorthogonal polarization, and argued that the PA trajectories of the 
polarization modes reflect the true viewing geometry of the pulsar. This 
interpretation preserves the RC model and avoids the unnecessary and unpalatable 
complication of invoking an additional radiation emission mechanism to explain 
the pulsar's modal connecting bridge. The occurrence of NPM in other pulsars is 
bound to cause their PA sweeps to deviate from the RC model, and although largely 
ignored in the past, NPM should be taken into account when determining pulsar 
viewing geometries in future work. The continuing ability of the statistical 
model to simulate the rich variety in the observed polarization properties of 
pulsars lends support to the model's applicability and its fundamental, underlying 
assumption that the polarization modes occur simultaneously. 

\acknowledgements
I thank Dan Stinebring for providing the data used in this analysis, and Joanna
Rankin for constructive comments on the manuscript. The Arecibo Observatory is
operated by Cornell University under cooperative agreement with the National 
Science Foundation.


\clearpage

\begin{references}
\reference{} Allen, M. C. \& Melrose, D. B. 1982, Proc. Astron. Soc. 
             Aust., 4, 365

\reference{} Backer, D. C. 1973, \apj, 182, 245

\reference{} Backer, D. C. \& Rankin, J. M. 1980, \apjs, 42, 143 (BR)

\reference{} Backer, D. C., Rankin, J. M., \& Campbell, D. B. 1976, 
             \nat, 263, 202 (BRC)

\reference{} Barnard, J. J. \& Arons, J. 1986, \apj, 302, 138

\reference{} Blaskiewicz, M., Cordes, J. M., \& Wasserman, I. 1991, 
             \apj, 370, 643

\reference{} Cheng, A. F. \& Ruderman, M. A. 1979, \apj, 229, 348

\reference{} Cordes, J. M., Rankin, J. M., \& Backer, D. C. 1978, \apj,
             223, 961

\reference{} Deshpande, A. A. \& Rankin, J. M. 2001, \mnras, 322, 438

\reference{} Drake, F. D. \& Craft, H. D. 1968, \nat, 220, 231

\reference{} Everett, J. E. \& Weisberg, J. M. 2001, \apj, 553, 341

\reference{} Gangadhara, R. T. 1997, \aap, 327, 155

\reference{} Hankins, T. H. \& Rickett, B. J. 1986, \apj, 311, 684

\reference{} Hankins, T. H., Rankin, J. M., Stinebring. D. R., \&
             McKinnon, M. M. 1992, in Proceedings of IAU Coll 128, The 
             Magnetospheric Structure and Emission Mechanisms of Radio 
             Pulsars, ed.  T. Hankins, J. Rankin, \& J. Gil, (Zielona 
             G\'ora: Pedagogical University Press), 161

\reference{} Hibschman, J. A. \& Arons, J. 2001, \apj, 546, 382

\reference{} Manchester, R. N., Taylor, J. H., \& Huguenin, G. R. 
             1975, \apj, 196, 83 (MTH)

\reference{} Mardia, K. V. 1972, Statistics of Directional Data,
             (London: Academic), 24

\reference{} McKinnon, M. M. 2002, \apj, 568, 302

\reference{} McKinnon, M. M. \& Stinebring, D. R. 1998, \apj, 502, 883
             (MS1)

\reference{} McKinnon, M. M. \& Stinebring, D. R. 2000, \apj, 529, 435
             (MS2)

\reference{} Michel, F. C. 1987, \apj, 322, 822

\reference{} Radhakrishnan, V. \& Cooke, D. J. 1969, Astrophys. Lett., 3, 
             225 (RC)

\reference{} Ramachandran, R., Rankin, J. M., Stappers, B. W., Kouwenhoven,
             M. L. A., \& van Leeuwen, A. G. J. 2002, \aap, 381, 993

\reference{} Rankin, J. M. 1983, \apj, 274, 333

\reference{} Rankin, J. M. 1986, \apj, 301, 901

\reference{} Rankin, J. M. 1993, \apjs, 85, 145

\reference{} Rankin, J. M. \& Ramachandran, R. 2003, \apj, in press

\reference{} Stinebring, D. R., Cordes, J. M., Rankin, J. M., Weisberg, J. M., 
             \& Boriakoff, V. 1984, \apjs, 55, 247 (SCRWB)

\reference{} Taylor, J. H., Huguenin, G. R., Hirsch, R. M., \& Manchester,
             R. N. 1971, Astrophys. Lett., 9, 205

\end{references}
\end{document}